\documentclass[%
 reprint,
 amsmath,amssymb,
 aps,
]{revtex4-2}

\usepackage{graphicx}
\usepackage{dcolumn}
\usepackage{bm}

\usepackage{microtype}
\usepackage{xcolor}

\usepackage{siunitx}
\usepackage{ulem}

\usepackage[breaklinks=true]{hyperref}
\usepackage{breakcites}
\usepackage[nameinlink]{cleveref}
\usepackage{microtype}

\newcommand*{\vertbar}{\rule[-1ex]{0.5pt}{2.5ex}}
\newcommand*{\horzbar}{\rule[.5ex]{2.5ex}{0.5pt}}

\definecolor{darkblue}{rgb}{0.0,0.0,0.4}
\definecolor{darkgreen}{rgb}{0.0,0.25,0.0}
\definecolor{shadecolor}{gray}{1}
\definecolor{felix}{rgb}{0.09,0.65,0.53}

\hypersetup{
	colorlinks,
	linkcolor=darkblue,
	citecolor=darkblue,
	urlcolor=darkblue
}

\begin{document}

\title{Bose Einstein condensate as nonlinear block of a Machine Learning pipeline}

\author{Maurus Hans}
\affiliation{Kirchhoff-Institut für Physik, Im Neuenheimer Feld 227, Universität Heidelberg}
\author{Elinor Kath}
\email{physical\_computing\_1@matterwave.de}
\affiliation{Kirchhoff-Institut für Physik, Im Neuenheimer Feld 227, Universität Heidelberg}
\author{Marius Sparn}
\affiliation{Kirchhoff-Institut für Physik, Im Neuenheimer Feld 227, Universität Heidelberg}
\author{Nikolas Liebster}
\affiliation{Kirchhoff-Institut für Physik, Im Neuenheimer Feld 227, Universität Heidelberg}
\author{Felix Draxler}
\affiliation{Computer Vision and Learning Lab, Universität Heidelberg}
\affiliation{Image and Pattern Analysis Group, Universität Heidelberg}
\author{Christoph Schnörr}
\affiliation{Image and Pattern Analysis Group, Universität Heidelberg}
\author{Helmut Strobel}
\affiliation{Kirchhoff-Institut für Physik, Im Neuenheimer Feld 227, Universität Heidelberg}
\author{Markus K. Oberthaler}
\affiliation{Kirchhoff-Institut für Physik, Im Neuenheimer Feld 227, Universität Heidelberg}

\date{\today}

\begin{abstract}
Physical systems can be used as an information processing substrate and with that extend traditional computing architectures. 
For such an application the experimental platform must guarantee pristine control of the initial state, the temporal evolution and readout. 
All these ingredients are provided by modern experimental realizations of atomic Bose Einstein condensates. 
By embedding the nonlinear evolution of a quantum gas in a Machine Learning pipeline, one can represent nonlinear functions while only linear operations on classical computing of the pipeline are necessary.
We demonstrate successful regression and interpolation of a nonlinear function using a quasi one-dimensional cloud of potassium atoms and characterize the performance of our system.
\end{abstract}

\maketitle

A very general approach to harness the computational resources of a system is to embed it in a Machine Learning pipeline, which processes data sequentially.
Each block in the pipeline must be able to compute functions of its input.
For example, continuous-time input can be processed by a dynamical system, such as a recurrent artificial neural network \cite{jaegerEchoStateApproach2001, natschlagerLiquidComputerNovel2002, verstraetenExperimentalUnificationReservoir2007, lukoseviciusReservoirComputingApproaches2009}. 
Physical systems can also compute expressive functions, making them strong candidates for a computational block in a Machine Learning pipeline, if they are reliable. This has been demonstrated previously \cite{nakajimaPhysicalReservoirComputing2020, tanakaRecentAdvancesPhysical2019, nakajimaReservoirComputingTheory2021}.

Only recently, it has been suggested that a Bose Einstein condensate (BEC) of ultracold atoms is an ideal system to be embedded in a Machine Learning pipeline \cite{marcucciTheoryNeuromorphicComputing2020, silvaReservoirComputingSolitons2021}: 
It allows for precise control, and its complex nonlinear dynamics depend unambiguously on the input.
The dynamics of the order parameter of a BEC are well described by a nonlinear differential equation (Gross-Pitaevskii equation).
Thus, it can be employed as a nonlinear physical layer of a Machine Learning pipeline. 
We address the concrete task of computing the value $u = f(x)$ of a given nonlinear function $f$, employing a standard computer for linear operations only. 
Thus we perform a regression, which is another application of such models besides the common task of classification.
In the context of Machine Learning, this is known as the Linear Basis Function Model \cite[Chapter 3.1]{bishopPatternRecognitionMachine2006}. 

\begin{figure}
	\centering
	\includegraphics[width=\columnwidth]{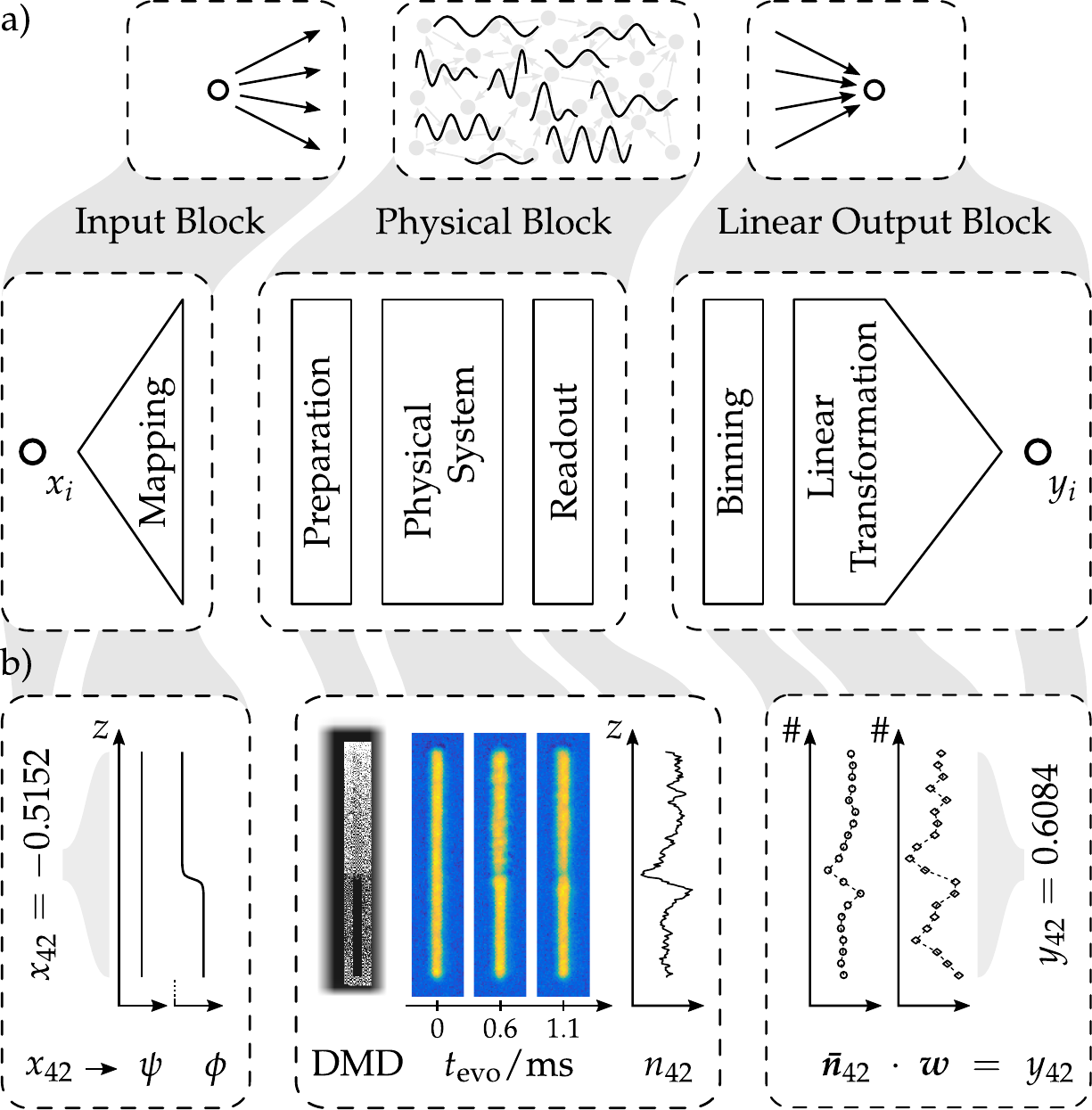}
	\caption{\textbf{Machine learning pipeline.}
    (a) In the input block, the input $x$ is mapped to a degree of freedom of a physical system. 
    The physical block includes the actual preparation of the corresponding initial state, a nonlinear time evolution and the readout. 
    Finally, in the output block, the physical readout is binned and a trained linear transformation produces the output. 
    (b) An input value $x_i$ is mapped to the phase function $\phi(z)$ of a quasi one-dimensional Bose Einstein condensate. 
    The phase imprint is realized using a digital micromirror device (DMD). 
    After the time evolution the density distribution is read out. 
    The density is further processed by binning and transformed via a vector multiplication to a single output value.}
	\label{Fig1}
\end{figure}

\paragraph*{The model}
The structure of the overall scheme is depicted in \cref{Fig1}a and consists of an input block, a physical block with a Bose Einstein condensate (BEC) as the central element, and a linear output block, where the latter is defined by supervised learning (see \cref{Fig1}b). 
In the first, linear block, one of $N_x$ input values $x_i$ is mapped onto the mean-field of freedom of a BEC, namely the complex order parameter $\Psi= \psi e^{i\phi}$, with amplitude $\psi(z)$ and phase $\phi(z)$. 
As the representation of the number $x_i$ we choose a step in the phase at a certain spatial position $z_i$.
The physical block is defined by the imprinting of the phase, a subsequent evolution, and a readout of the density.
The resulting one-dimensional density $n_i(z)=|\Psi_i(z)|^2$ corresponds to the input $x_i$.
The linear output block consists of two layers: 
The first improves signal-to-noise by binning the density $n_i(z)$ to $N_w$ bins, leading to a data vector ${\bf \bar{n}_i} = (n_{i,1},...,n_{i,N_w})$. 
The second layer performs a linear transformation $y_i = {\bf \bar{n}_i} \cdot {\bf w}$, where the weight vector ${\bf w} = (w_1,...,w_{N_w})$ is a result of supervised learning.
In the example given in \cref{Fig1}b, the function $f(x) = \sin(\pi x)/(\pi x)$ is implemented, and for the input value of $x_{42}=-0.5152$ we find $y_{42} = 0.6084$, close to $u_{42} = f(-0.5152) = 0.6172$. 

\paragraph*{Implementation}
Our physical system is a quasi one-dimensional BEC of ${}^{39}\mathrm{K}$ atoms with a length of $85\,\mu$m and a width of $6\,\mu$m confined in a box trap. 
The condensate is confined in the gravitational direction with a trap frequency of $1.5\,$kHz, realized with a blue-detuned ($532\,$nm) lattice with a spacing of $\sim5\,\mu$m.
In the other direction a digital micromirror device (DMD) is used to realize the quasi one-dimensional configuration, employing laser light at the same wavelength.
The phase step encoding the input is experimentally implemented by shining the blue-detuned light on a specific area of the atomic cloud for $100\,\mu$s inducing an estimated phase shift of $1.4\,\pi$ on one side of the cloud (see \cref{Fig1}b).
Using the DMD, we imprint the phase step at arbitrary positions on the atomic cloud with excellent reproducibility. 
Careful adjustment of the light intensity for the box trap avoids changes of the trapping frequency along the short axis during the imprint.
After one input value is encoded, the atomic cloud evolves freely in the box trap. 
The imprint leads to the formation of a density peak, which moves along the cloud and disperses, as well as a stationary but decaying density dip (see \cref{Fig3}a).
The positions of the phase imprint are in the center-half of the condensate minimizing the effect of the edges.
The density profile is extracted after a variable evolution time $t_{evo}$ via two-frequency absorption imaging \cite{hansHighSignalNoise2021}. 
The resulting image of the elongated density distribution is integrated along the short axis of the cloud.

\paragraph*{Training the linear output block} 
In the first layer of the linear output block, the values of the extracted density are binned leading to a vector ${\bf \bar{n}_{i}}$ of length $N_{w}$.
Then, the final linear transformation is defined by supervised learning, minimizing the root-mean-square (RMS)  error
\begin{equation}
    \varepsilon = \sqrt{\frac{1}{N} \sum\limits_{i=1}^{N} ( y_i - u_i )^2} \; .
    \label{Eq:RMSE}
\end{equation}
where $y_i \in \mathbb{R}$ are the results of the model and $u_i = f(x_i)\in \mathbb R$ are the prescribed function values to be learned.
Since $N_w < N_x$, this is achieved by calculating the Moore-Penrose pseudo-inverse \cite{penroseBestApproximateSolutions1956} for the system of linear equations
\begin{equation}
    \begin{pmatrix}
		\  \bar{n}_{1,1} & \hdots & \bar{n}_{1,N_w} \\
		\vdots  & & \vdots\\
		\  \bar{n}_{N_x,1} & \hdots & \bar{n}_{N_x,N_w} \\
	\end{pmatrix}
    \begin{pmatrix}
		w_1 \\ \vdots \\ w_{N_w}
	\end{pmatrix}
	=
	\begin{pmatrix}
		u_1 \\ \vdots \\ u_{N_x}
	\end{pmatrix}
    \:,
    \label{Eq:LinSystem}
\end{equation}
constructed from measured $\bf \bar{n}_i$ and target values $u_i$ associated with the input values $x_i$. 
Note that the number of bins is also the number of weights $N_w$.
Therefore, there is an optimal $N_w$ much smaller than $N_x$, where high signal-to-noise as well as expressive power is given.

\begin{figure}
	\centering
	\includegraphics[width=\columnwidth]{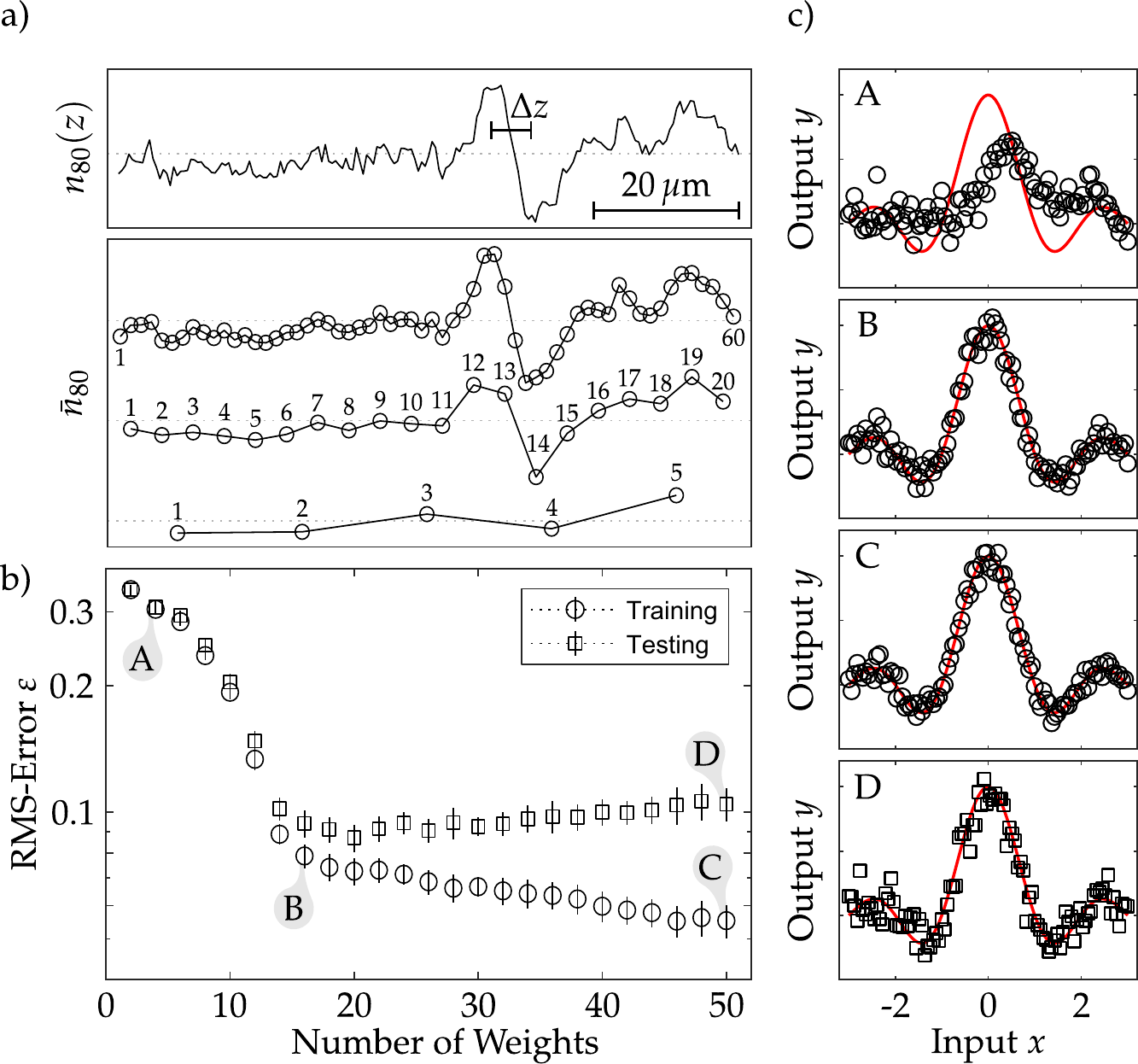}
	\caption{\textbf{Performance dependence on number of weights.} 
    (a) Density distribution for an evolution time of $1.1\,$ms after the phase imprint (top).
    Lowering the number of bins $N_w$ reduces the spatial resolution and leads to less information about the original distribution, while reducing the noise (bottom).
    (b) RMS-error of training and testing as a function of $N_w$.
    For training the RMS-error decreases for higher bin numbers, with a characteristic kink at $\sim18$ bins.
    This behaviour is caused by the dependence of both expressivity and signal-to-noise on the number of bins.
    Error bars have been calculated by bootstrap resampling analysis (see methods).
    (c) Examples for training (A-C) and testing (D) for different number of bins. The red lines indicate the target functions.}
	\label{Fig2}
\end{figure}

\begin{figure}
	\centering
	\includegraphics[width=\columnwidth]{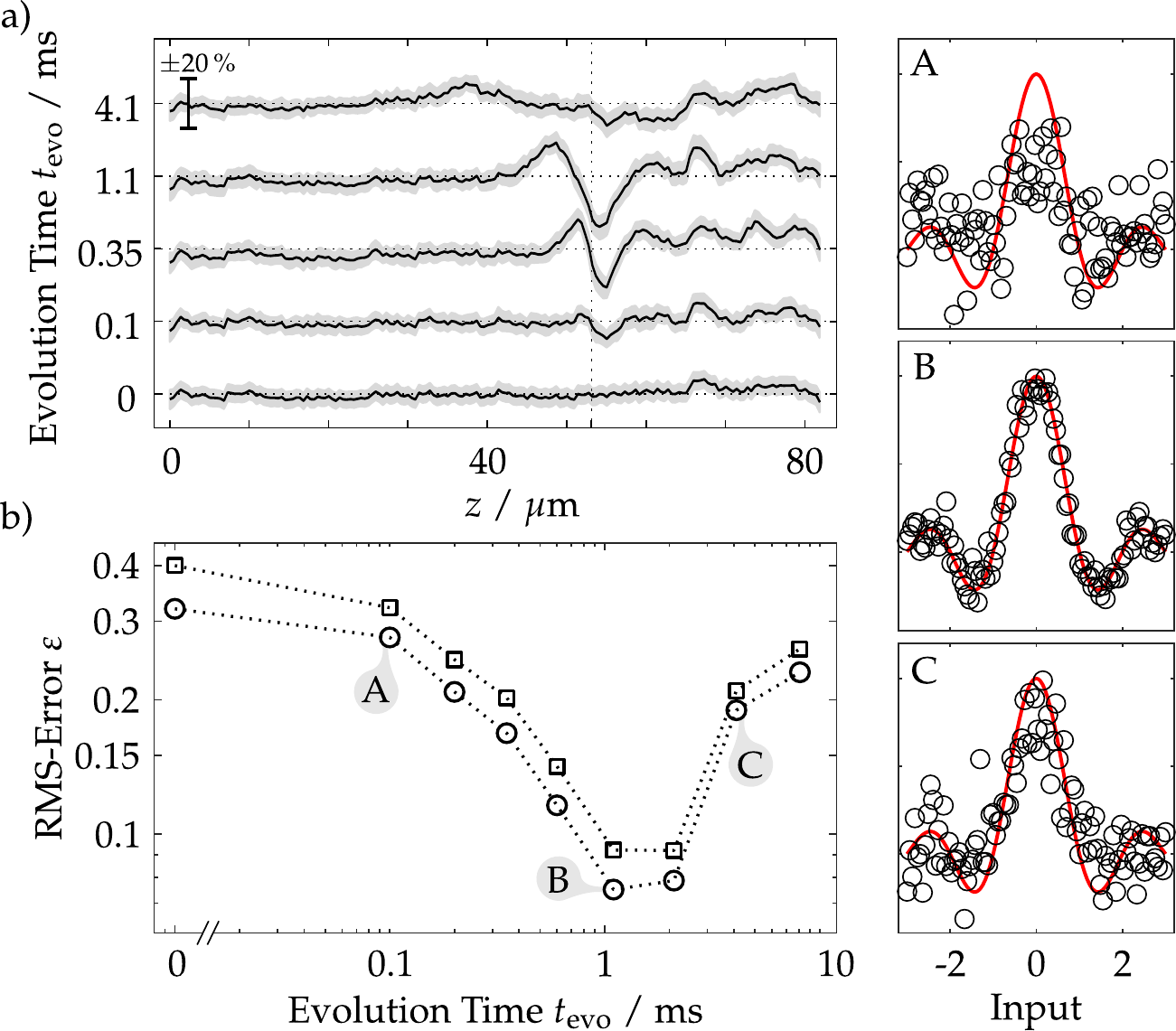}
	\caption{\textbf{Performance dependence on evolution time.} 
    (a) Density distributions after different evolution times. 
    Depicted is the mean over 30 realizations, with the standard deviation given by the shaded area.
    Over time the imprint leads to a density peak, which moves to the left and widens, and a non-moving density dip. 
    For evolution times longer than $2\,$ms both features slowly decay.
    (b) RMS-error of training and testing as a function of evolution time. 
    For each evolution time the number of weights $N_w$ is chosen such that the RMS-error of testing is minimal.
    An optimum for both training and testing can be found at $1.1\,$ms, where the signal-to-noise of the binned density profiles is maximal.}
	\label{Fig3}
\end{figure}

\paragraph*{Regression Task.}
We solve a regression task for the function $f(x) = \sin(\pi x) /(\pi x)$, a standard nonlinear function for benchmarking. 
For that purpose we compile the dataset $\left\{ (x_i,u_i) \right\}$ from $N_x = 100$ evenly spaced input values $x_i \in [-3,3]$ with their associated target values $u_i = f(x_i)$. 
The experiment is run for each input value $x_i$ with the associated phase imprint and an evolution time of $t_\mathrm{evo} = 1.1\,\mathrm{ms}$. 
An example of the resulting density distribution for the specific input value $x_{80}$ is shown in the upper trace of \cref{Fig2}a.  
The linear transformation is then trained by inverting \cref{Eq:LinSystem} with binned density profiles $\bf \bar{n}_i$ (see \cref{Fig2}a, lower), resulting in an optimized weight vector $\bf w$. 
This training is successful if the weight vector leads to viable results not only for training data, but also for new experimental realizations of the same input values (testing).

Training on differently binned data reveals that there is an optimal binning giving the best performance as can be seen in \cref{Fig2}b. 
The RMS-error for training monotonically decreases with the number of weights $N_w$, and ultimately vanishes as the matrix in \cref{Eq:LinSystem} reaches full rank, {\it i.e.} the number of inputs equals the numbers of weights.
A characteristic kink at $\sim 18$ weights can be identified.
While the RMS-error of the training suggests better performance with a further increase of $N_w$, testing clearly reveals that the performance of the regression does not improve beyond the kink but saturates and ultimately decreases. 
As can be seen in \cref{Fig2}a the number of bins $N_w$ affects the spatial resolution.
Thus the system's expressivity is limited not only by the number of weights $N_w$, but also by the information extracted from the physical system.
A minimal sampling is required to resolve the density feature, and therefore the substantial output information is accessed given $N_w > 2 Z/\Delta z=\tilde{N}_{w}$, where $\Delta z$ is the characteristic extension of the density feature and $Z$ is the length of the system used for information encoding (see Methods), in close analogy to the Nyquist-Shannon sampling criterion \cite{shannonCommunicationPresenceNoise1949}.

The deviation of the performance of training and testing occurs because the training becomes dominated by a generic property of a physical system - the noise. 
For ultracold gases this is ultimately given by the shot noise due to the finite number of particles.
With binning this noise level is reduced since adjacent pixels are averaged.
Because fewer pixels are averaged as the number of bins increases, the noise on the density profiles increases as well.
This increase of noise can be partially compensated in the training data due to the increasing number of weights, \textit{i. e.} expressivity; however, when applying the trained weights to testing data, the noise directly translates into poorer performance.
As a consequence the best performance of a physical layer strongly depends on $N_w$, which has to be optimized to balance information extraction and noise reduction. 
This is system specific and has to be determined for a chosen physical system. In Machine Learning, this phenomenon is known as the bias-variance trade-off or decomposition \cite[Chapter 3.2]{bishopPatternRecognitionMachine2006}.

\paragraph*{Time evolution of the physical system}
To analyse the relevance of the physical evolution, the quality of the regression is investigated for different evolution times. 
\Cref{Fig3}a shows the averaged (30 realizations) density profiles corresponding to the input value $x_{80}$ for different evolution times. 
The profiles share the same scale and are shifted by a constant offset for clarity. 
The standard deviation of the mean is given by the grey shaded area. 
After the preparation a peak of expelled atoms moves to the left while the depletion remains stationary.
This scheme of phase imprinting has been employed for dark soliton preparation \cite{burgerDarkSolitonsBoseEinstein1999, denschlagGeneratingSolitonsPhase2000} and the observed phenomenology suggests that the emerging dark soliton decays due to a snake instability expected for the given trap geometry \cite{cornellSnakeInstability2001}.

For a quantitative analysis the model is trained with different evolution times of the physical system. 
The resulting RMS-error $\varepsilon$ with $N_w$ chosen optimally at each evolution time is shown in \cref{Fig3}b for training and testing. 
With increasing evolution time the error decreases and reaches a minimum at $t_\text{evo}=1.1\,$ms. 
This behaviour can be understood from the evolution of the profiles, where the density features resulting from the phase imprint have to develop first and widen.
For times shorter than $1.1\,$ms the narrow density features imply that the optimal number of bins $\tilde{N}_w\propto Z/\Delta z$ is high. 
Thus the resulting signal-to-noise ratio is low, limiting the performance.
The increase in RMS-error for significantly longer times is mainly a result of the decreasing signal due to the spreading of the peak and filling of the depletion. 
The testing error is close to the training error for all times, demonstrating successful learning (in contrast to overfitting, see Methods).
The significant reduction of the error for evolution times on the order of $1\,$ms indicates a high predictive power of our model.

\begin{figure}
	\centering
    \vspace{0.3cm}
	\includegraphics[width=\columnwidth]{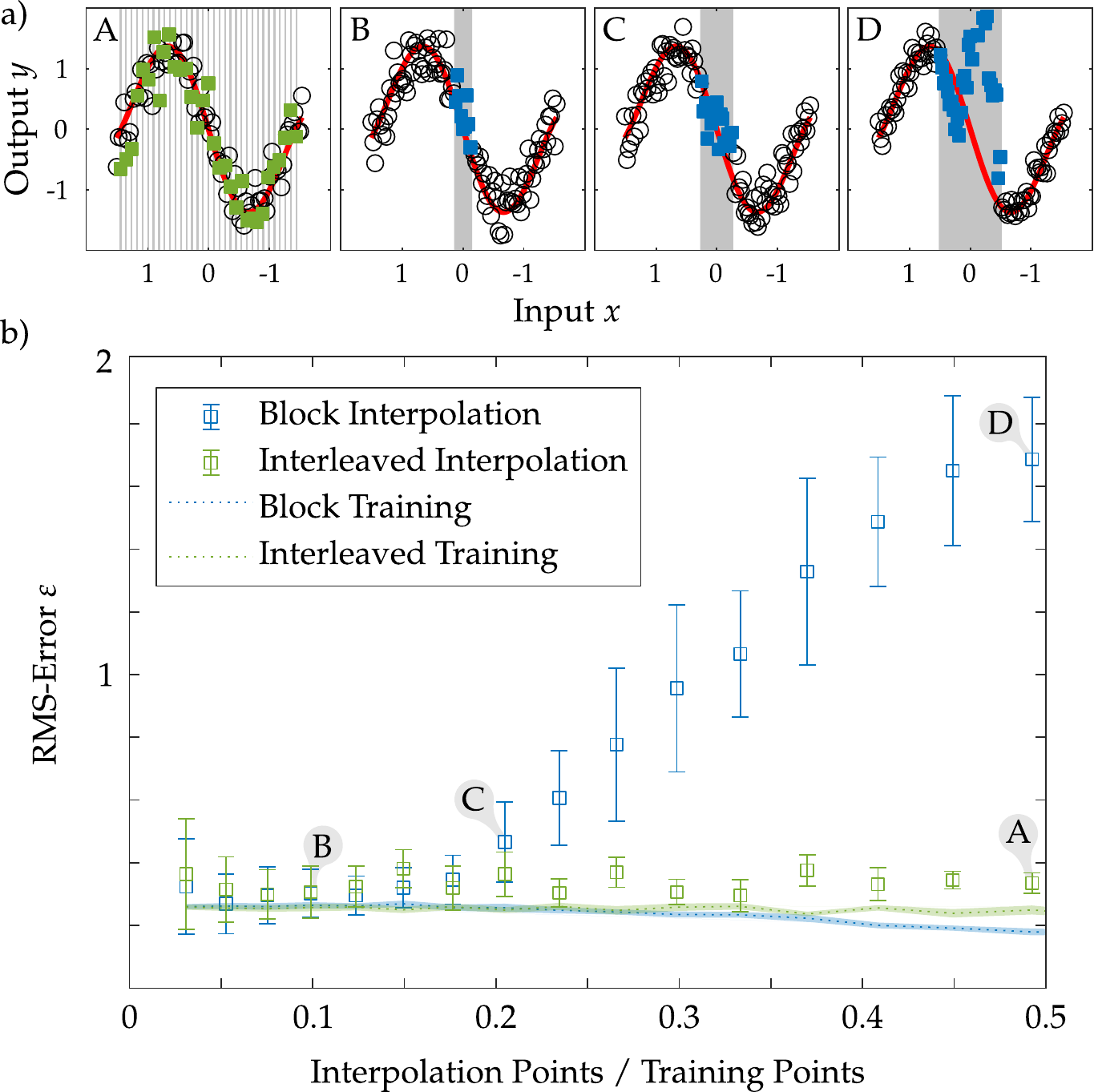}
	\caption{\textbf{Generalization.} 
    The model is used for interpolation between trained input values. 
    (a) Examples of the performance. Red line: Target function, circles: training points, squares: interpolation points.           
    The interpolation inputs are either equally spaced over the whole interval (interleaved case), or all from the central region (block case). 
    The interpolation regions are indicated by grey shading.
    The interpolation between adjacent training points (interleaved) works very well (A) but fails beyond a critical block length (B-D).  
    (b) Dependency of the RMS-error on the ratio between interpolation and training points. 
    The interleaved case works well for all ratios. 
    For ratios beyond $0.2$, corresponding to a block size of $17$ interpolation points, the performance drastically decreases which can be understood as the failure of the corresponding wavelet decomposition (details see text).
    Error bars are the result of bootstrap resampling analysis.}
	\label{Fig4}
\end{figure}

\paragraph*{Generalization}
To estimate the model's ability of generalization we test the performance with input values $x_i$ that are not part of the training data set. 
We define good performance of a regression task as the robust and reliable interpolation between the given supporting points $x_i$. 
For the following discussion we choose the derivative of the former target function $f(x) = \sin(\pi x)/(\pi x)$, since it is better suited to reveal the direct connection between our Machine Learning model and Linear Basis Function Models \cite[Chapter 3.1]{bishopPatternRecognitionMachine2006}. 
We use the physical block at its best performance given by $t_{\text{evo}}=1.1\,$ms and binning of $N_w=18$, and take the $N_x=100$ inputs equally spaced in the range of $x_i\in[-1.5, 1.5]$.
We investigate the performance of the interpolation both for equally spaced interpolation points over the whole input range (interleaved) as well as for interpolation points in a single domain (block, see \cref{Fig4}a).
Panel (A) shows the good performance in interpolating every third input qualitatively.
This is confirmed by the RMS-error (see \cref{Fig4}b), which hardly depends on the distance between interpolated points.
Block interpolation, however, works only up to a certain block size (panels (B)-(D)).
The RMS-error grows beyond a ratio between interpolation points and training points of $0.2$, corresponding to block sizes larger than $\Delta x = 17$ input points.
This behaviour can be understood by reinterpreting the data matrix in \cref{Eq:LinSystem} to
\begin{equation}
    	\begin{pmatrix}
		\ \horzbar\!\!\! & \bar{n}_1 & \!\!\!\horzbar\ \ \\
		& \vdots & \\
		\ \horzbar\!\!\! & \bar{n}_{N_x} & \!\!\!\horzbar\ \ \\
	\end{pmatrix}
\equiv \begin{pmatrix}
		\ \vertbar\!\!\!  &        & \!\!\!\vertbar\ \ \\
		  \  g_1         & \hdots &   g_{N_w}       \\
		\ \vertbar\!\!\!  &        & \!\!\!\vertbar\ \ \\
	\end{pmatrix},
 \label{eq:Representation}
\end{equation}
where the introduced $(g_j)_i$ is the mean density in bin $j$ given input $x_i$. 
From this perspective the linear transformation in the output block composes the discretized function $f(x_i) = u_i$ from the limited number of $(g_j)_i$ \cite[Chapter 4]{Hardt:2022uz}. 
This is the discretized version of a decomposition of the target function into a finite set of representative (basis) functions 
\begin{equation}
f(x) = \sum_{j=1}^{N_w} w_j\, g_j(x),
\label{Eq:wavelet}
\end{equation}
which is known as wavelet analysis and at the heart of Linear Basis Function Models. 
Our physical system produces basis functions of spatially localized dispersive signals $g_j(x)$.
Due to the diagonal nature of our data matrix (see Methods), the relative width of the basis function $\Delta x / X$, where $X$ is the range of $x$ values, is similar to the relative width of the density features in $\bf\bar{n}_i$, $\Delta z / Z$.
Interpolation does not work if the basis function corresponding to the center of the interpolation region has vanishing overlap with the basis functions corresponding to the edges of the block. 
This explains the failure of interpolation beyond extended regions of $N_x\Delta z /Z \sim 17$ interpolation points, which is equivalent to a ratio of interpolation to training points of 0.2 (see \cref{Fig4}b). 

\paragraph*{Conclusion and Outlook}
The demonstration of a Machine Learning model with a quasi one-dimensional highly controlled Bose Einstein condensate as the nonlinear physical block opens the general perspective of fusing highly controlled physical systems with classical computing. 
The approach and results are agnostic to the underlying theoretical description of the ongoing physics but rely on precise experimental control. 
We have chosen the regression problem for the first demonstration since the performance can be straight forwardly quantified. 
It also clearly demonstrates the parallels between our Machine Learning model and the paradigmatic Linear Basis Function Model. 
Since our physical system offers a broad range of precisely adjustable parameters (plasticity), such as nonlinearity (interaction strength), potential landscapes and temperature, the input block as well as the physical block could be optimized for specific tasks.  
Especially the perspective of utilizing quantum features of the physical system in Machine Learning architectures opens up new possibilities for information processing.

We thank Celia Viermann and Thomas Gasenzer for fruitful discussions. This work is supported by the Deutsche Forschungsgemeinschaft (DFG, German Research Foundation) under Germany's Excellence Strategy EXC2181/1-390900948 (the Heidelberg STRUCTURES Excellence Cluster), and within the Collaborative Research Center SFB1225 (ISOQUANT, Project-ID 273811115). N. L. acknowledges support by the Studienstiftung des deutschen Volkes.

\newpage

\normalem
\bibliography{references_phd}

\clearpage
\section*{Supplementary}
\paragraph*{Experimental System.} 
We prepare a Bose Einstein condensate of ${}^{39}\mathrm{K}$ with approximately $20{,}000$ atoms in the substate corresponding to $F=1, m_F=-1$ at low magnetic fields. 
The scattering length is tuned to $50\,a_\mathrm{B}$ ($a_\mathrm{B}$ being the Bohr radius) by applying a homogeneous magnetic field to exploit the Feshbach resonance at $561\,\mathrm{G}$ \cite{etrychPinpointingFeshbachResonances2022, derricoFeshbachResonancesUltracold2007}. 
An additional magnetic gradient levitates the atoms against gravity. 
In vertical direction the atomic cloud is confined in a single lattice site of a repulsive lattice created by blue-detuned laser beam ($532\,\mathrm{nm}$ light, lattice spacing $5\,\mu\mathrm{m}$). 
This leads to a strong two-dimensional confinement with a trap frequency of $\omega \sim 2\pi \times 1.5\,\mathrm{kHz}$. 
In the horizontal plane a configurable dipole potential is applied with a blue-detuned $532\,\mathrm{nm}$ laser beam, which is shaped by a digital micromirror device (DMD) in direct imaging configuration \cite{gauthierDirectImagingDigitalmicromirror2016}. 
To avoid uncontrolled interferences between vertical and horizontal confinement, the light frequencies are shifted by $160\,$MHz.
The DMD also allows to imprint phase shifts on the atoms by illuminating one area of the cloud.
The density distribution of the atomic cloud is extracted by absorption imaging with a two-frequency scheme\cite{hansHighSignalNoise2021}. 
The resolution for both setups is $\sim 1\,\mu\mathrm{m}$. 
We estimate the chemical potential from the velocity of the density peaks to be $\sim 1.2\,\mathrm{kHz}$ and the healing length to $\sim 0.3\,\mathrm{\mu m}$.\\

\paragraph*{Bootstrap Resampling Analysis.}
In order to estimate the statistical deviation of the RMS-error with a limited number of experimental runs, we employ a method inspired by bootstrap resampling \cite{efronIntroductionBootstrap1994}. 
For every input value $x_i$ (where $i = 1\dots N$) we measure $N_\text{R}$ different density profiles $n_{i,k}\,$. 
We then compile a set of density profiles $\mathcal{N}_\text{Train}$ by randomly choosing one of the realizations $k_i = 1\dots N_\text{R}$ for every input value $x_i\,$,
\begin{equation}
	\mathcal{N}_\text{Train} = \left\{ n_{1,k_1} \dots \, n_{N,k_{N}} \right\} \;.
\end{equation}
For this set of experimental realizations the training is applied to obtain the weight vector $\textbf{w}$.
We then calculate the RMS-error $\varepsilon$ for the training set according to \cref{Eq:RMSE} to judge the quality of the regression. 
Next, a testing set $\mathcal{N}_\text{Test}$ is put together the same way as the training set $\mathcal{N}_\text{Train}$, but using only density profiles $n_{i,k} \notin \mathcal{N}_\text{Train}\,$. 
Then, the RMS-error $\varepsilon$ is calculated for the testing set $\mathcal{N}_\text{Test}$. 
This technique allows the generation of many different training sets from a limited number of experimental realizations per input value. 
The results for the RMS-error shown in this publication are calculated from the mean and standard deviation of 25 repetitions of this process.\\

\begin{figure}
	\centering
    \vspace{0.4cm}
	\includegraphics[width=\columnwidth]{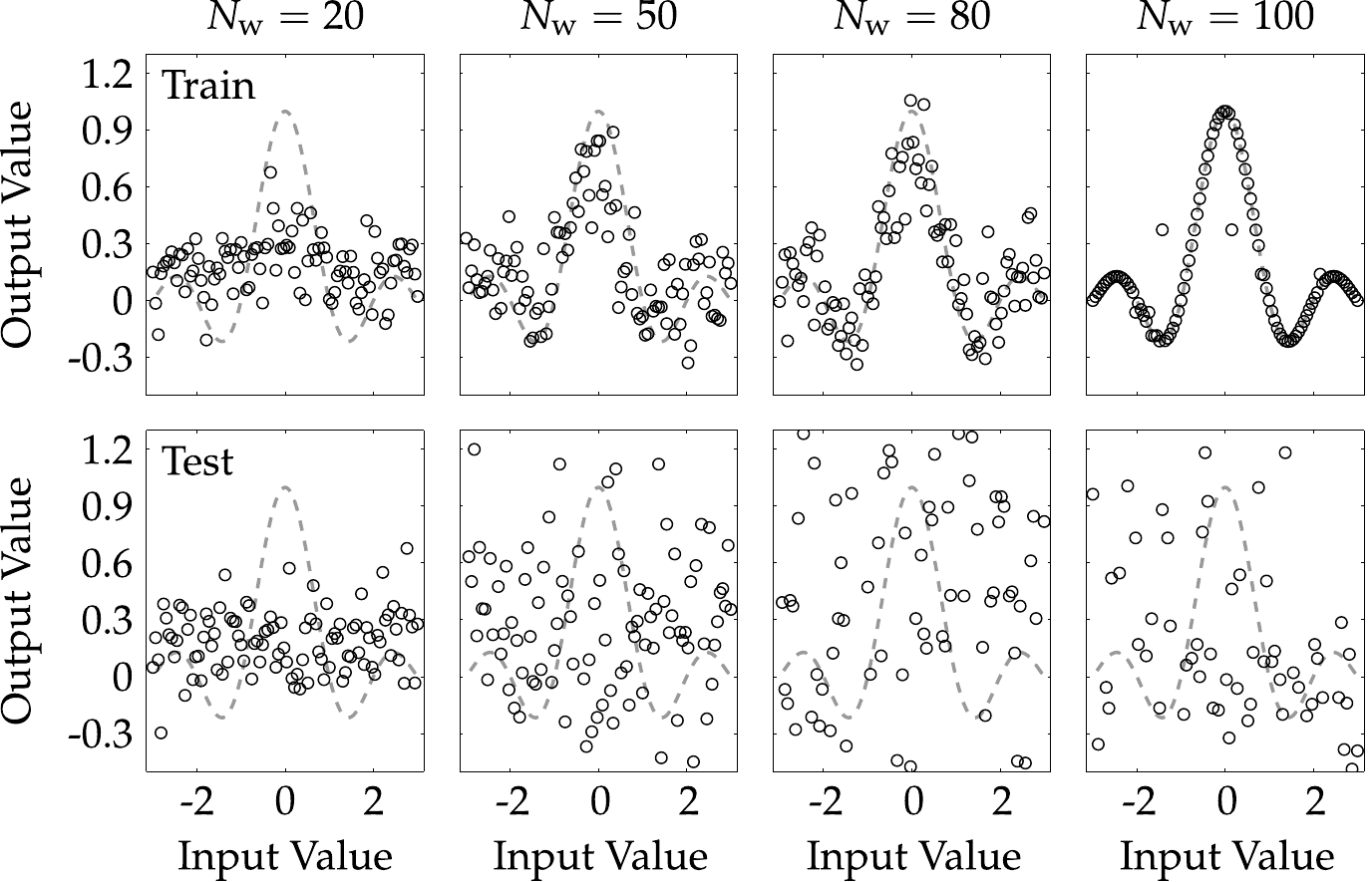}
	\caption{\textbf{Example of overfitting.} The top four plots show the result of training a regression, where reference profiles were used for each input value ($N=100$). They are completely uncorrelated to the input as they are obtained without the phase imprint. However, by increasing the number weights $N_w$ a regression to the target function (dashed line) seems possible. Testing the obtained weight vector $w$ with a different set of inputs reveals that no prediction can be made and the results stem from overfitting (lower row).}
	\label{Fig:Overfitting}
\end{figure}

\paragraph*{Overfitting.}
To illustrate the importance of testing the obtained weight vector $\textbf{w}$, we sabotage the training procedure by using reference profiles as the input profiles $n_i$. 
These noisy but flat profiles do not have any phase imprinted and are thus uncorrelated to input values $x_i$. 
However, by increasing the number of weights $N_w$ a fit becomes possible (\cref{Fig:Overfitting}, upper row). 
If the number of parameters matches the number of training points and all input profiles are linearly independent, \cref{Eq:LinSystem} becomes invertible and the training data is matched. 
Since there is no information encoded in the data, testing with data previously unseen by the model will fail. 
In the lower row of \cref{Fig:Overfitting} this lack of predictive power is obvious.\\

\begin{figure}
	\centering
    \vspace{0.5cm}
	\includegraphics[width=\columnwidth]{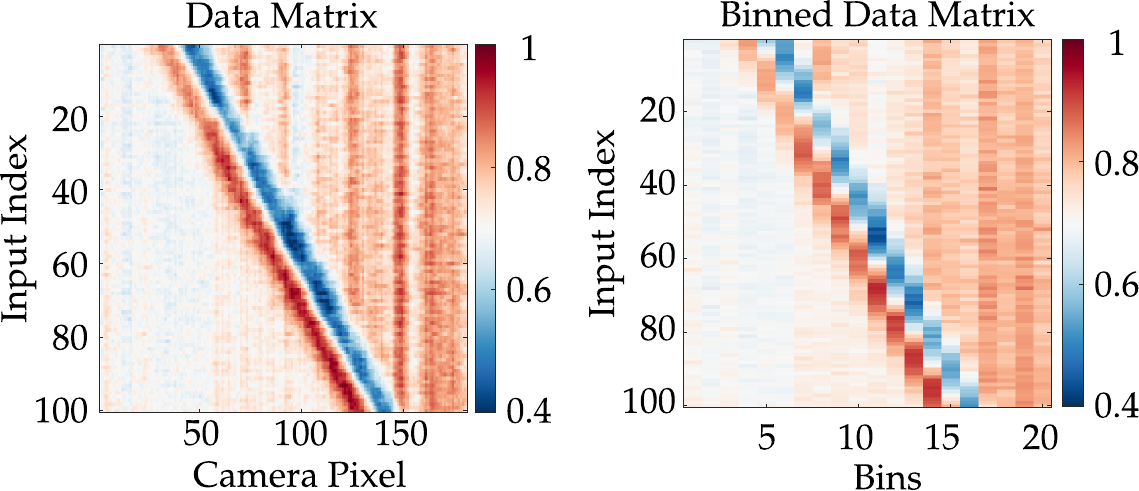}
	\caption{\textbf{Data Matrix.} Left: Bare data. Right: Binned data.}
	\label{Fig:DataMatrix}
\end{figure}

\paragraph*{Data matrix.}
The one-dimensional data vectors for all input values and $t_\mathrm{evo} = 1.1\,\mathrm{ms}$ are shown in \cref{Fig:DataMatrix}, in the left plot.
Each row is the detected one-dimensional density $n_i$ on the camera for input value $x_i$.
One sees the emergent density peak and dip.
The position of the pattern is linearly dependent on the input value, and thus $\Delta x /X \propto \Delta z/Z$.
On the right side of the phase step, small artifacts in the density are produced when the DMD is switched on to imprint a phase on one portion of the cloud. 
The right plot shows the data set binned for optimal performance, \textit{i. e.} 20 bins.

\end{document}